\documentclass[conference,a4paper]{IEEEtran}

\IEEEoverridecommandlockouts
\usepackage{cite}
\usepackage{amsmath,amssymb,amsfonts}
\usepackage{algorithmic}
\usepackage{graphicx}
\usepackage{textcomp}
\usepackage{xcolor}

\usepackage{comment}
\usepackage{algorithm}
\usepackage{algorithmic}
\usepackage{breqn}
\usepackage{graphicx}
\usepackage{subcaption}
\usepackage{graphicx}
\usepackage{footnote}
\usepackage[bottom]{footmisc}
\makesavenoteenv{table}
\makesavenoteenv{tabular}
\usepackage{threeparttable}
\usepackage{tabularx}
\usepackage{ragged2e}
\newcolumntype{L}[1]{>{\raggedright\arraybackslash}p{#1}}
\newcolumntype{C}[1]{>{\centering\arraybackslash}p{#1}}
\newcolumntype{R}[1]{>{\raggedleft\arraybackslash}p{#1}}
\newcolumntype{J}[1]{>{\justifying\arraybackslash}p{#1}}
\usepackage{booktabs}

\makeatletter
\newcommand\fs@norules{\def\@fs@cfont{\bfseries}\let\@fs@capt\floatc@ruled
  \def\@fs@pre{}%
  \def\@fs@post{}%
  \def\@fs@mid{\kern3pt}%
  \let\@fs@iftopcapt\iftrue}
\makeatother

\usepackage{amssymb}

\usepackage{soul}
\usepackage{xcolor}
\usepackage{xparse}

\usepackage{amsmath}
\DeclareMathOperator*{\argmax}{argmax}

\usepackage[belowskip=-0.00pt,aboveskip=0pt]{caption} 
\def\BibTeX{{\rm B\kern-.05em{\sc i\kern-.025em b}\kern-.08em
    T\kern-.1667em\lower.7ex\hbox{E}\kern-.125emX}}
    
\begin{document}

\title{
Edge-Aided Sensor Data Sharing\\in Vehicular Communication Networks\thanks{This work was supported by the German Federal Ministry of Transport and Digital Infrastructure (BMVI) in the project “KIVI -- KI im Verkehr Ingolstadt” and the Bavarian Ministry of Economic Affairs, Regional Development and Energy (StMWi) in the project ``IN2Lab
-- Ingolstadt Innovation Lab''.}
}

\author{
\IEEEauthorblockN{Rui Song\IEEEauthorrefmark{1}\IEEEauthorrefmark{2},
Anupama Hegde\IEEEauthorrefmark{3},
Numan Senel\IEEEauthorrefmark{4},
Alois Knoll\IEEEauthorrefmark{2},
Andreas Festag\IEEEauthorrefmark{1}\IEEEauthorrefmark{3}}
\IEEEauthorblockA{\IEEEauthorrefmark{1}Fraunhofer Institute for Transportation and Infrastructure Systems IVI\\
Ingolstadt, Germany, Email: \{rui.song $|$ andreas.festag\}@ivi.fraunhofer.de}
\IEEEauthorblockA{\IEEEauthorrefmark{2}Technical University of Munich, Chair of Robotics, Artificial Intelligence and Real-time Systems\\
Garching, Germany, Email: rui.song@tum.de, knoll@in.tum.de}
\IEEEauthorblockA{\IEEEauthorrefmark{3}Technische Hochschule Ingolstadt, CARISSMA Institute for Electric, COnnected, and Secure Mobility (C-ECOS)\\
Ingolstadt, Germany, Email: \{anupama.hegde $|$ andreas.festag \}@carissma.eu}
\IEEEauthorblockA{\IEEEauthorrefmark{4}Technische Hochschule Ingolstadt, 
Institute of Innovative Mobility (IIMo)\\
Ingolstadt, Germany, Email: numan.senel@thi.de}}

\maketitle

\begin{abstract}
Sensor data sharing in vehicular networks can significantly improve the 
range and accuracy of environmental perception for connected automated vehicles. %
Different concepts and schemes for dissemination and fusion of sensor data have been developed. 
It is common to these schemes that measurement errors of the sensors impair the perception quality and can result in road traffic accidents. %
Specifically, when the measurement error from the sensors -- also referred as measurement noise~-- is unknown and time varying, the performance of the data fusion process is restricted, which represents a major challenge in the calibration of sensors. 
In this paper, we consider sensor data sharing and fusion in a vehicular network with both, vehicle-to-infrastructure and vehicle-to-vehicle communication.
We propose a method, named \emph{Bidirectional Feedback Noise Estimation} (BiFNoE), in which an edge server collects and caches sensor measurement data from vehicles. The edge estimates the noise and the targets alternately in double dynamic sliding time windows and enhances the distributed cooperative environment sensing at each vehicle with low communication costs. 
We evaluate the proposed algorithm and data dissemination strategy in an application scenario by simulation and show that the perception accuracy is on average improved by around 80\,\% with only 12\,kbps uplink and 28\,kbps downlink bandwidth. 
\end{abstract}

\begin{IEEEkeywords}
sensor data sharing and fusion, cooperative perception, vehicular communication, distributed estimation, measurement noise
\end{IEEEkeywords}

\section{Introduction}

Sensor data sharing enables connected automated vehicles (CAV) to acquire 
environmental information by exchanging their locally perceived sensor data via communication among themselves and with the roadside infrastructure. It enhances the range and the accuracy of the environmental perception and thereby improves the safety and efficiency in road traffic. 
In order to reduce the data load in the resource-constrained communication network, the perceived sensor data are rather shared as lists of detected and classified objects than as raw data~\cite{Guenther2016_CP}.
Subsequently, the sensor data fusion is realized on object level, either in a distributed or a centralized fashion~\cite{HE202021}.

Sensor fusion based on shared object lists in vehicular communication networks is able to provide better perception.
However, the sensor measurement quality in CAVs is characterized by measurement noise, always unknown and time-varying, which leads to failures in the sensor data fusion results. %
This paper studies a novel approach for cooperative estimation of sensor measurement noise in vehicular environments purely based on sensor data generated by CAVs. %
Other research on this topic, e.g.,~\cite{10.1115/DSCC2017-5075}, proposed an online noise identification method for a single sensor that has not taken the information sharing in the (sensor) network into account. %
Although the concept of transmitting the quality of sensor data as part of the messages exchanged among vehicles is being considered (see e.g. in~\cite{C2CCC}), the measurement noise in each CAV can change 
due to effects from road and weather conditions, individual deterioration of hardware, diverse combinations of sensor types, bias in training data, etc. Also, it might not be feasible to share the information about inaccuracy of CAVs because of the data sensitivity and privacy reasons.

\begin{figure}[t]
   \centering
   \includegraphics[width=0.35\textwidth]{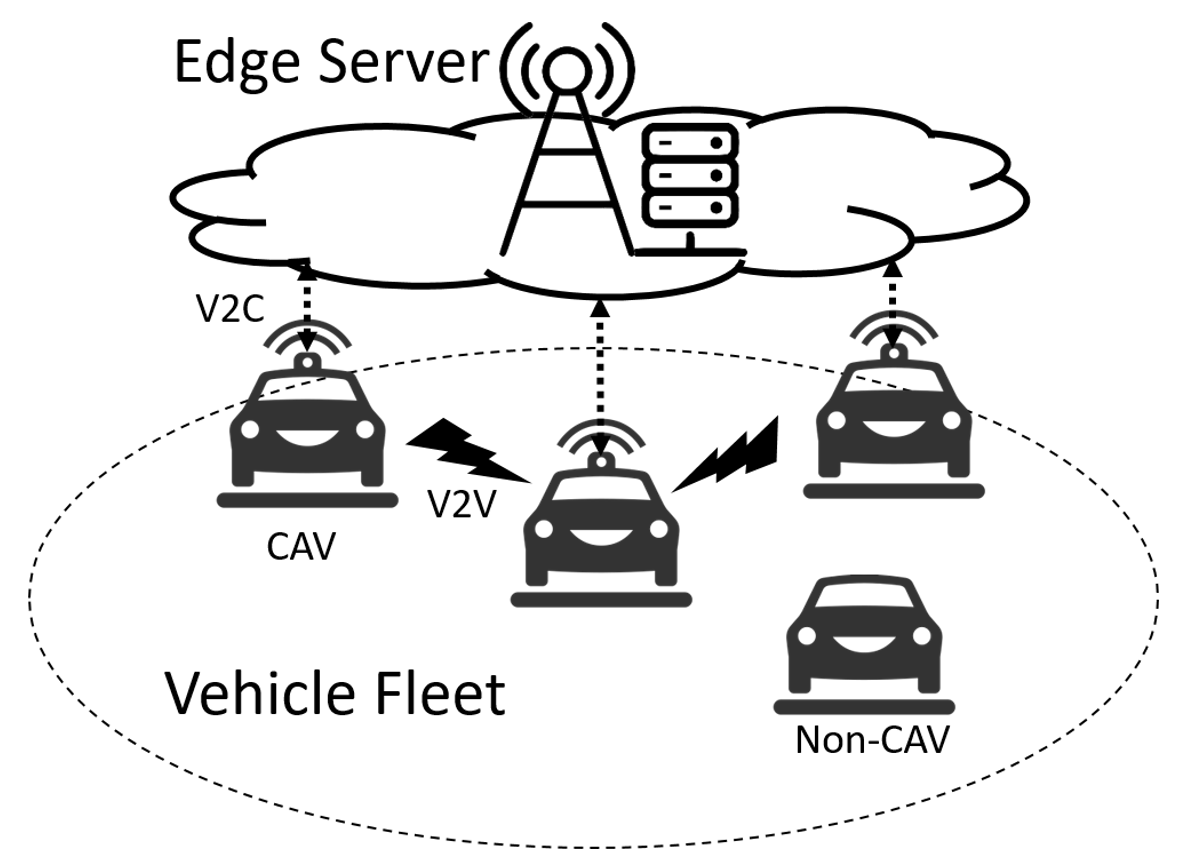}
   \caption{Overview of the sensor data sharing system in vehicular networks, relying on vehicle-to-cloud (V2C) and vehicle-to-vehicle (V2V) communication between an edge server and an associated vehicle fleet, including CAVs and non-CAVs.}
   \label{fig:system}
\end{figure}

The system architecture considered in this paper is shown in Fig.~\ref{fig:system}: CAVs can share sensor data acquired by on-board sensors with an edge server by means of vehicle-to-cloud (V2C) communications or among each other by vehicle-to-vehicle (V2V) communications. 
In the first case, the edge server performs centralized sensor data fusion. In the latter case, distributed data fusion is applied in every individual vehicle. We can assume that centralized data fusion provides better perception results compared to distributed data fusion in each CAV, the transmission of the fusion results from the edge to the vehicles requires latency in millisecond level that is hard to achieve. On the other side, the sensor measurement noise in distributed sensor data fusion can cause heavy errors. We believe that the noise estimation for distributed data fusion 
is necessary and essential.

In this paper, we propose to estimate the sensor measurement noise in each CAV at a computation center located at the edge of the network close to the vehicles. This edge server is capable of sharing the noise knowledge to all CAVs in a vehicle fleet and to enhance the distributed object-level sensor data fusion. The key contributions of the paper are:

\begin{itemize}
    \item Formulating the problem of unknown noise values in sensor data sharing based on vehicular communication,
    \item Developing an 
    algorithm (Bidirectional Feedback  Noise Estimation, BiFNoE) that determines the noise estimations for each CAV and improves the centralized target estimation at an edge server,
    \item Proposing a communication service for noise estimation enhancing the distributed estimation at a single CAV,
    \item Providing a thorough evaluation and investigating the trade-off between accuracy and communication costs, %
    \item Implementing the proposed system with the BiFNoE algorithm in a simulation environment and evaluating 
    the proposed approach in an urban scenario. 
\end{itemize}

The remainder of this paper is structured as follows. Section~\ref{sec:related_work} summarizes related work in sensor data sharing fusion in vehicular network. The problem with unknown noise values in a multi-agent system is formulated in Section~\ref{sec:problem_formulation}. The algorithm for noise estimation is introduced in Section~\ref{sec:algorithm} and the associated system is proposed in Section~\ref{sec:protocol}. The performance of algorithm and protocol is comprehensively evaluated in Section~\ref{sec:evaluation} considering communication costs.  Section~\ref{sec:conclusion} provides a summary and concludes.

 \section{Related Work}
\label{sec:related_work}

Environmental perception is commonly regarded as a major challenge for the realization of vehicle automation. Sensor data sharing among vehicles helps increasing the range and the accuracy of perception. 
Previous work on sensor data sharing has primarily studied the data dissemination in the context of vehicular communication for ad hoc networks, see 
e.g.,~\cite{Delooz2020,Garlich2019,Thandavarayan2020,schiegg2021}, and for vehicle-to-cloud communication~\cite{FISITA2021}. Sensor data sharing is also being considered in the standardization processes of vehicular communication~\cite{ETSI-TR103562-CPS-studyitem,SAE_CP}. To acquire a better estimation of the states of a target vehicle, e.g., its position and velocity, based on sensor data sharing, 
\cite{8916837,Shorinwa2020,zhang2021distributed} investigated data fusion methods.

Among the wide spectrum of existing work in the area of sensor data sharing and fusion, the distributed estimation method is a key approach, especially for the fusion of shared data at object level. %
By applying consensus-based Kalman Filter methods~\cite{Olfati2005,5399678,BATTISTELLI2016169},
a CAV is able to aggregate the perceived information from on-board sensors and from all other CAVs in the communication range, thereby improving the state estimation of the target vehicles. Several distributed estimation strategies and protocols have been proposed, 
e.g., in~\cite{5411741,yuan2017distributed,SHIN2020259,9204471}, where only a finite number of measurements from reachable nodes is taken into account. In~\cite{HE202021}, these publications were summarized and compared by simulations in terms of global optimality, local consistency, communication overhead 
 and topology requirements.

In vehicular applications and others, the sensors' measurement noise in each CAVs, which is unknown and time-varying, can cause a bad estimation. The existing methods for handling the noise  aim either at the use of advanced statistical models~\cite{9195416,8036249,8326722}
or at a better calibration process as part of the manufacturing process. Both methods are not suitable for the sensor data fusion in vehicular scenarios. The reasons are:
\begin{itemize}
    \item 
    The measurement providers are only temporary reachable CAVs in the vehicular network.
    \item 
    The measurement inaccuracy depends on the environmental conditions, e.g. road, weather, lighting, etc.
    \item 
    The performance of AI-based measurements, i.e., detection results, can be affected by the bias in training data.
\end{itemize}
While state-of-the-art for automatic calibration using camera~\cite{6909022,10.1145/3199667}, LiDAR~\cite{9010134,8317829} or hybrid sensor combinations~\cite{8917135} can provide the knowledge about inaccuracy of the on board sensors' detection a vehicle, but requires an additional advanced sensor set-up in roadside units. Moreover, the potential detection failures in roadside units can affect the quality for each CAV implementing the automatic calibration. 

In this paper, we address those challenges in vehicular applications and employ a novel approach to the problem of noise estimation. We apply the 
estimation in continuously available edge servers and CAVs in a dedicated vehicle fleet, where the CAVs can upload their sensor data to the edge server. Compared to the automatic calibration approaches, our proposed solution does not require an additional sensor setup. Instead, we optimize the noise estimation in double dynamic sliding time windows at the edge server. 
By subscribing the proposed noise estimation service, the CAVs improve their estimations, which converge to the limit.
 
\section{Problem Formulation}
\label{sec:problem_formulation}

As we consider a vehicle fleet at time $t$ with a number of CAVs with index $i$ and normal vehicles in scenarios of partial connected traffic, $N_t$ and $M_t$ are the number of CAVs and normal vehicles. The detected targets with index $l$ for each CAV are included in $\mathcal{D}_{i,t} = \{z_{i,l,t}|l=1,2,...,L_{i,t}\}$, where $L_{i,t}$ is the the number of objects detected by CAV $i$ at time $t$ and ideally $L_{i,t}\leq N_t+M_t-1$. Meanwhile, we consider the edge server as one of the connected agents with index $c$ in the communication network, which can acquire all CAVs' perception information $\mathcal{D}_{c,t}=\{\mathcal{D}_{i,t}|i=1,2,...,N\}$. When we model the global communication network at time $t$ as a graph $\mathcal{G}_t = (\mathcal{V}, \mathcal{E}_t)$, for CAV $i$, the set of connected vehicles is $\mathcal{N}_{i,t} = \{j|(i,j)\in\mathcal{E}_t\}$. Note that for the edge server, we yield $\mathcal{N}_{c,t} = \{j|(c,j)\in\mathcal{E}_t\} = \{1,2,...,N\}$.

In an vehicular ad hoc network, a CAV broadcasts its own perception information using a standardized message type~\cite{ETSI-TR103562-CPS-studyitem}, which transmits lists of detected objects and other information. For each detected object, i.e., a target vehicle $l$, its state transition can be modeled as a system function $f$, then
\begin{equation}
\label{Eq:state_transition}
x_{l,t}=f_{l}(x_{l,t-1})+w_{l,t}
\end{equation}
where $x_{l,t}$ corresponds to the states of target vehicle $l$ at time point $t$, and $w_l$ is the noise value of system transition. If the perception process of CAV $i$ modeled as $h_i$, the perception result of target vehicle $l$ is 
\begin{equation}
    \label{Eq:measurement}
    z_{i,l,t}=h_{i}(x_{l,t})+v_{i,t}
\end{equation}
where $v_{i,t}$ is the measurement noise of CAV $i$. In the communication network, one target can be redundantly perceived multiple times by different CAVs. This can be described in one agent $i$ as $\{z_{i,l,t}, z_{j,l,t}|j\in \mathcal{N}_{i,t}\}$ for a single target vehicle~$l$. Based on the cooperative perception information and the system transition model, the final perception results $\hat x_{i,l,t}$ can be generated with (\ref{Eq:state_transition}) and (\ref{Eq:measurement}) by a distributed estimation model:
\begin{equation}
\label{Eq:est}
\hat x_{i,l,t} = \argmax_{x_{l,t}} p (x_{l,t}|\bar{x}_{l,t}, z_{i,l,t}, z_{j,l,t}|j\in \mathcal{N}_{i,t})
\end{equation}

In (\ref{Eq:est}), each target $l$ is maintained by each CAV $i$ as an independent track. After track association using the Hungarian algorithm~\cite{Bewley2016}, the local track $z_{i,l,t}$ and $\{z_{j,l,t}|j\in \mathcal{N}_{i,t}\}$ from the neighbors are weighted and aggregated. The prior of mean $\bar{x}_{l,t}$ and the state $\hat x_{i,l,t}$ in posterior can be estimated  based on the aggregation of independent perceptions with the knowledge of the measurement noise $v_i$, which is necessary but actually unknown. 
Presuming the measurement noise $v_i$ follows a certain distribution, the relation between observed target $z_{i,l,t}$ and ground truth $x_{l,t}$ is
\begin{equation}
z_{i,l,t}=g(\Theta_i, h(x_{l,t}))
\end{equation}
where $\Theta_i$ is the parameter matrix of the noise model and can be estimated by Maximum-Likelihood Estimation (MLE). The 
results are especially trustworthy, when a large number of targets are observed by 
multi-independent observers. 

As the 
edge server can communicate to the CAVs in a vehicle fleet, 
each CAV node is able to upload its sensor data to the central backend. Through centralized estimation, the edge server can generate the environment sensing results $\{\hat x_{c,l,\tau}|l=[1:L_{c,t}], \tau=[0:t]\}$, which are more accurate than the distributed 
estimation results at each CAV node. If these results can be generated in real time and downloaded by each CAV, the environmental perception results for all CAVs can be as good as in the central backend. However, this can hardly be achieved because of the limited bandwidth and the inherent communication latency. Consequently, the inaccurate noise values will decrease the quality of the sensor data fusion and eventually pose a threat to the traffic safety.

\section{Proposed Noise Estimation Algorithm}
\label{sec:algorithm}

The aim of sensor data sharing in each CAV $i$ is estimating the states of each target $l$ at time point $t$, i.e., $x_{l,t}\in \mathbb{R}^{n}$. When the measurement noise of CAV $i$, $v_i$, is modeled as a Gaussian distribution, i.e., $v_i \sim  \mathcal{N} (0, R_i)\in \mathbb{R}^{n}$, the noise model parameter matrix $\Theta_i$ can be represented as covariance matrix denoted by $R_i$. By executing Distributed Kalman Filter (DKF), each CAV is able to estimate the states of target vehicles detected by on-board sensors and neighbor CAVs. Normally, the measurement noise values at most CAVs in one vehicle fleet are relative small and nearly constant within a time scale of minutes, i.e., for most $i \in \mathcal{N}_{c,t}$: $v_{i,t} = v_{i,t-T}$. Then we assume for most $l\in \mathcal[1:L_{i,t}]$ and most $t\in [0:t-1]$
\begin{equation}
\label{eq:z}
z_{i,l,t} \sim  \mathcal{N} (h_{i}(x_{l,t}), R_i),
\end{equation}
where $L_{i,t}$ is the maximum enumerated index of targets in one locally detected object list at CAV $i$ at time point $t$.

To estimate the measurement noise value at each CAV, we propose a \emph{Bidirectional Feedback in Double Time Windows} (BiFNoE) algorithm that dig the available perception information in an edge server and alternately improve the noise and target estimations in time series. The estimated noise values can be updated and transmitted periodically. In this way, they enhance the results of the distributed estimation at each CAV.

\begin{figure}[t]
   \centering
   \includegraphics[width=0.5\textwidth]{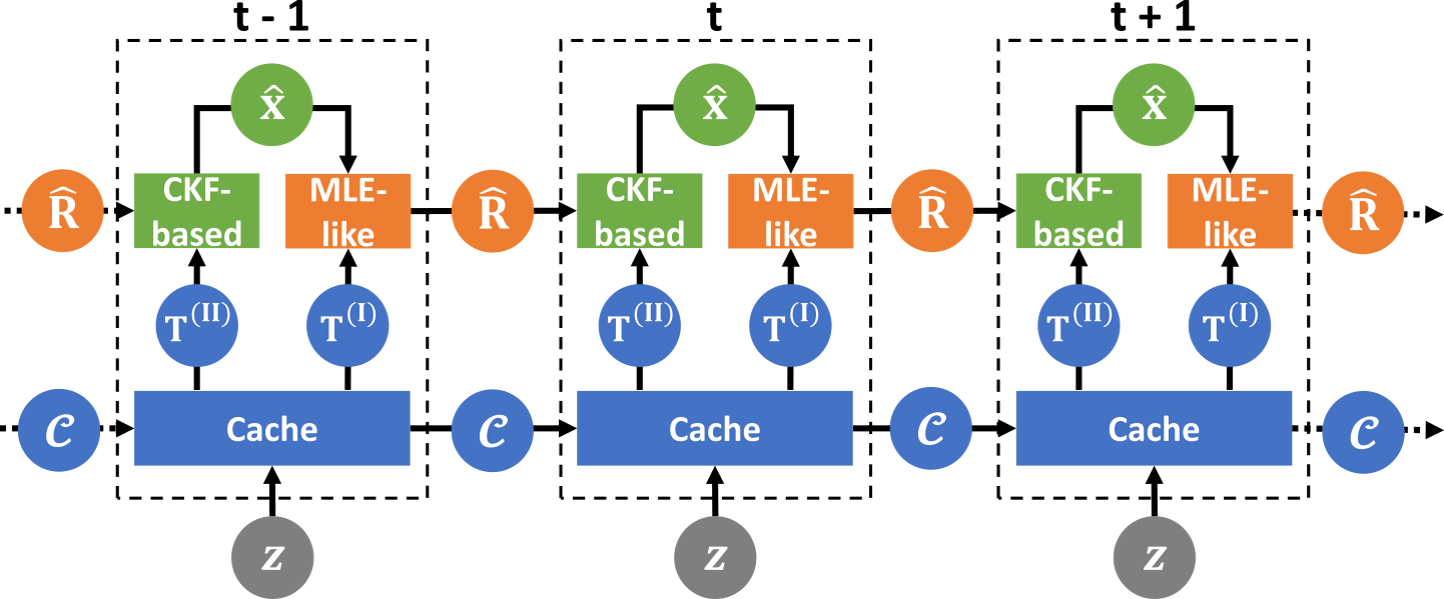}
   \caption{Bidirectional feedback between CKF-based and MLE-like operations in double sliding time windows $T^{(I)}$ and $T^{(II)}$.}
   \label{fig:bidrectional}
\end{figure}

We assume that the Centralized Kalman Filter (CKF) is employed at edge server. The noise values for all CAVs compose a covariance matrix $R_c$ of CKF, which can be initialized as $\hat{R}_{c,0}$ for the target estimation. With an increasing number of the measurement samples, the noise features of each CAV becomes prominent. The $\hat{R}_{c}$ can be then computed by MLE with the target estimations, i.e., an MLE-like method, based on more promising results at the edge server:
\begin{multline}
\label{eq:R_est}
\hat{R}_{c} =\argmax_{R_{c}} p(
z_{i\in \mathcal{N}_c,l \in [1:L_t],t\in [0:T]},\\
\bar{x}_{i\in \mathcal{N}_c,l \in [1:L_t],t\in [0:t-1]}
|R_c, 
\hat{x}_{i\in \mathcal{N}_c,l \in [1:L_t],t\in [0:t]})
\end{multline}
The noise value is characterized as the concentration of the observed samples from a time period $t\in[0:t]$ at various detected targets $l \in [1:L_t]$ with respect to the central estimated results $\hat{x}_{i\in \mathcal{N}_c,l \in [1:L_t],t\in [0:t]}$. The multi-target tracking in a multi-agent system enables the enlargement of the sample number. In order to additionally enrich the data resource, especially for a vehicle fleet with low traffic density, the first sliding time window $[t-T^{(\rm I)}:t]$ is proposed. Then the target estimation term in (\ref{eq:R_est}) is replaced by $\hat{x}_{i\in \mathcal{N}_c,l \in [1:L_t],t\in [t-T^{(\rm I)}:t]}$. By increasing $T^{(\rm I)}$, the sample number can be controlled over a predefined threshold, which guarantees the performance of noise estimation even with small $L_t$ and $\mathcal{N}_c$. The estimated noise values enhance the centralized target estimation, which further increase the noise estimation quality in time series. The process of bidirectional feedback between CKF-based and MLE-like operations in double silding time windows is illustrated in Fig.~\ref{fig:bidrectional} and the BiFNoE pseudo-code in Algorithm~\ref{alg:BiFNoE}.

 \begin{algorithm}[t]
 \caption{\raggedright Proposed BiFNoE algorithm} 
 \label{alg:BiFNoE}
 \begin{algorithmic}[1]
 \renewcommand{\algorithmicrequire}{\textbf{Input:}}
 \renewcommand{\algorithmicensure}{\textbf{Output:}}
 
 \REQUIRE {{$\hat{R}_{c,t-1}$, $\{\mathcal{D}_{i,t}|i \in \mathcal{N}_{c,t}\}$, $\hat{x}_{c,t-1-T^{(\rm II)}:t-1}$}, $P_{c,t-1}$}
 \ENSURE  {$\{\hat{R}_{i,t}|i \in \mathcal{N}_{c,t}\}$}
    \\ \textit{Implementing Target Estimation in Time Window II}:
    \STATE Update $T^{(\rm II)}$ based on deviation of $P_{c,t-T^{(\rm II)}:t}$
\FOR {$i \in \mathcal{N}_{c,t}$}
    \STATE Update $\mathcal{C}_{c, t-T^{(\rm II)}:t}$ with $z_{i,t-T^{(\rm II)}:t}$
  \ENDFOR
 \FOR {$\mathcal{D}_{\tau} \in \mathcal{C}_{c, t-T^{(\rm II)}:t}$} 
  \STATE $(\hat{x}_{c,\tau}^{*}, P_{c,\tau}^{*})\gets$ Equation (\ref{Eq:est})
  \ENDFOR
  \STATE $P_{c,t}\gets P_{c,t}^*$
  \STATE $\hat{x}_{c,t}\gets \hat{x}_{c,t}^*$
      \\ \textit{Implementing Noise Estimation in Time Window I}:
  
  \FOR {$i \in \mathcal{N}_{c,t}$}
  \STATE Update $T^{(\rm I)}$ based on $\Sigma_{\tau=t-T^{(\rm I)}}^{t}{L_\tau}$ 

\STATE {$\hat{R}_i\gets$ MLE-like method with $\hat{x}_{c,t}$ and Equation (\ref{eq:z})}
  \ENDFOR

 \end{algorithmic}
 \end{algorithm}

\section{Application Scenario} 
\label{sec:protocol}

\begin{figure*}[ht]
\centering
\includegraphics[width=0.9\textwidth]{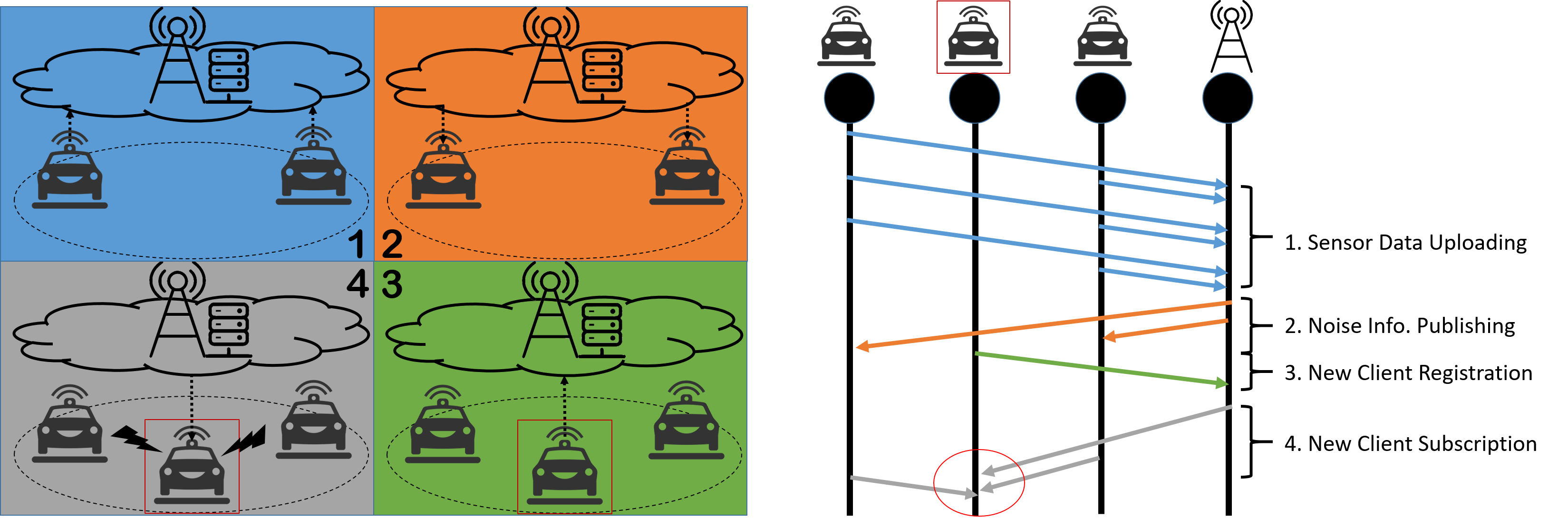}
\caption{Four stages (in four colors) of our proposed communication system for the noise estimation service}
\vspace{-0.3cm}
\label{fig:Protocol}
\end{figure*}

In order to integrate the algorithm in a vehicle fleet, we also take the distinct performance and properties between the vehicle and edge server into account and propose a communication system for noise estimation service as shown in Fig.~\ref{fig:Protocol}. The system is divided into four stages that are marked in different colors. At the stage of \emph{Sensor Data Uploading}, a vehicle fleet with a number of CAVs is considered, which upload their sensor data to an edge server, including at least a registered ID, timestamp and the object states detected by on-board sensors. 
By fusing the sensor data from CAVs using Algorithm~\ref{alg:BiFNoE}, the target states and perception noise values of the CAVs are centralized estimated. Instead of directly sharing the target states to each CAV, the edge server publishes the noise estimations at the stage of \emph{Noise Information Publishing} as a service that can be subscribed by all CAVs in the fleet. If a new CAV moves into this fleet and registers itself with the edge server at the third stage, \emph{New Client Registration}, the edge server receives and confirms the registration by creating a new vehicle ID for the new 
client. At the same time, this CAV can subscribe the noise value sharing service at the stage of \emph{New Client Subscription}, and then integrate the noise values in the distributed data fusion where the sensor data from neighbors are aggregated as marked in the red circle. The noise estimation service is described in Algorithm~\ref{alg:noise_service}.

\begin{algorithm}[t]
 \caption{Proposed noise estimation service}
 \label{alg:noise_service}
 \begin{algorithmic}[1]
  \renewcommand{\algorithmicrequire}{\textbf{Input:}}
 \renewcommand{\algorithmicensure}{\textbf{Output:}}
  \REQUIRE {$\hat{R}_{c,0}$ in the edge server}
  \ENSURE {$\hat{x}_{i,t}$ in each CAV agent}
    \\\textit {\textbf{Edge Server:}}
    \IF{Receive registration request}
    \STATE Assign registered ID for new CAV
    \ENDIF
    \FOR{Each CAV in ID list}
    \STATE {Update $\{\mathcal{D}_{i,t}|i \in \mathcal{N}_{c,t}\}$} from all registered CAVs
    \ENDFOR
    \STATE {Publish $\{\hat{R}_{i,t}|{i\in \mathcal{N}_{c,t}}\}$ by implementing BiFNoE}
    \\\textit {\textbf{Each CAV Agent i:}}
    \STATE {$\mathcal{D}_{i,t} \gets$ from on-board sensors}
    \IF{Available service not initialized}
    \STATE Send registration request
    \ENDIF
    \STATE {Upload the sensor data $\mathcal{D}_{i,t}$ to the edge server}
    \STATE {$\{\mathcal{D}_{j,t}|j \in \mathcal{N}_{i,t}\} \gets$ from other CAVs}
    \STATE {Broadcast the sensor data $\mathcal{D}_{i,t}$}
    \STATE {$\hat{R}_{c,t} \gets$ by subscribing}
    \STATE {$\hat{x}_{i,t} \gets $ DKF($\{\mathcal{D}_{j,t}|j \in \mathcal{N}_{i,t}\},\mathcal{D}_{i,t},\hat{R}_{c,t}$)}

 \end{algorithmic}
 \end{algorithm}

Note that the requirements of our system on the communication performance depend on \emph{(i)}~the sensor data upload frequency  $f_{upl}$, and \emph{(ii)}~the noise estimation service subscription frequency $f_{sub}$ in all CAVs. A high $f_{upl}$ assists the edge to update the environmental information faster and to reduce the age of information, which ensures noise value computing using BiFNoE with fresher sensor data. Also, a greater upload rate enables the edge to collect sufficient sensor data from each CAV within a shorter time window, especially for a new registered CAV. More freshness and fast growing volume of data is beneficial for the accuracy of the  BiFNoE results. Similarly, a higher $f_{sub}$ is helpful for CAV to download new updated noise information towards a better performance of perception with DKF. However, lower $f_{upl}$ and $f_{sub}$ are encouraged for saving the limited communication resources, in particular the uplink and downlink bandwidth for each CAV, because other edge services require dedicated quality of services as well. 
Therefore, appropriate $f_{upl}$ and $f_{sub}$ should guarantee the accuracy with respect to the requirements of related vehicle functions within relativly low communication costs.
Since the noise values of most CAVs in a vehicle fleet are nearly static, the BiFNoE and DKF do not require an 
extremely frequent information exchange between edge and CAV. Thereby the effectiveness of the noise estimation service can be guaranteed with compromised $f_{upl}$ and $f_{sub}$, which can reduce the communication overhead.
\section{Simulation Results}
\label{sec:evaluation}
\subsection{Algorithm Evaluation}

\begin{figure*}[ht]
\includegraphics[trim=0 0 0 0,clip,width=0.98\linewidth]{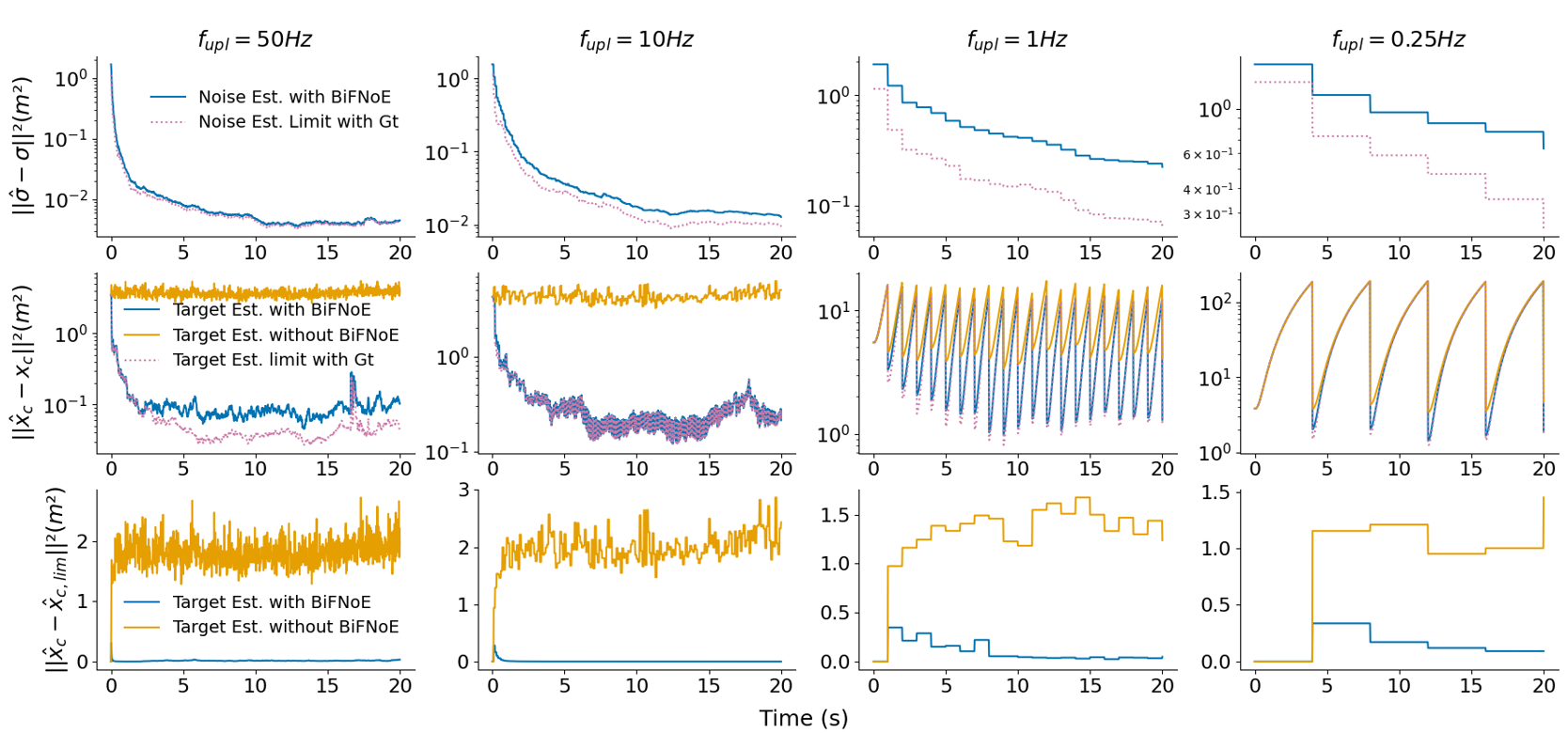}
\caption{Centralized noise and target estimation (est.) using BiFNoE. We observe the BiFNoE performance for different values of the sensor data uploading frequency, 50\,Hz, 10\,Hz, 1\,Hz and 0.25\,Hz in four columns. The centralized noise and target estimation accuracy is characterized by the mean squared error to ground truth (gt.) in the first two rows. We consider the noise estimation with target ground truth and target estimation with noise ground truth alternately as the limit. The last row shows the BiFNoE forced centralized target estimation to converge to the limit. The convergence speed is increased with a growing upload frequency of sensor data.}
\vspace{-0.4cm}
\label{fig:BiFNoE}
\end{figure*}

\noindent  \textbf{Experiment Design.} 
The algorithm is evaluated first in a simulation with N\,=\,100 CAVs and M\,=\,100 non-CAVs. Each CAV is able to detect the position of moving objects in its individual detection range, which is randomly generated within $[100:300]~m$. The diagonal of the covariance matrix $R_i$ for each vehicle is randomly set within $[0.01^2:5^2]$\,m\textsuperscript{2} in the initializing phase. Local perception results in form of object list $\mathcal{D}_{i,t}$ with registered ID and position are broadcasted to the neighbor CAVs within a communication range 150\,m and uploaded to the edge server. The simulation tick frequency $f_{sim}$\,=\,50\,Hz is the same as the broadcast frequency $f_{bdc}$ for data sharing in ad hoc network, while the upload frequency $f_{upl}$ is usually lower and can be varied.

\noindent  \textbf{Noise Estimation Analysis.} 
The centralized noise and target estimation using BiFNoE are shown in Fig.~\ref{fig:BiFNoE}. 
The first row depicts the MSE of the noise estimation with different sensor data upload frequency $f_{upl}$ with respect to the noise ground truth of each CAV. With an increasing iterative stage number in time series, the noise estimation approaches to the theoretical limit of the algorithm, where the ground truth of target $x_{c,l,\tau}$ instead of $\hat{x}_{c,l,\tau}$ is used as the mean of the noise model for noise estimation in time window~II (see Algorithm~\ref{alg:BiFNoE}). 

\noindent  \textbf{Centralized Estimation Analysis.} 
It is clear that the lower $f_{upl}$ reduces the convergence speed because less sensor data samples are collected for BiFNoE within the same number of time slots. However, the accuracy of the centralized target estimation can still be visibly increased compared with the estimation without BiFNoE, as shown in the second row in Fig.~\ref{fig:BiFNoE}. 
The bidirectional feedback enables the convergence of the centralized target estimation almost as fast as it in noise estimation. %
The MSE of the target estimation with respect to its limit is emphasized in the third row. It shows that the accuracy of the centralized target estimation can be notably increased and get close to the limit.

\subsection{System Evaluation}
\label{sec:system}

\noindent  \textbf{Experiment Design.} When the estimated noise values are transmitted as messages to each CAV in this vehicle fleet, the augmentation of the distributed estimation by subscribing the noise estimation service in different subscription frequencies $f_{sub}$ at edge is shown in  Tab.~\ref{table:dist_est}, where the $f_{sub}$ is $0$\,Hz, which means no subscription. The frequency of simulation $f_{sim}$ and sensor data uploading $f_{upl}$ are both $10$\,Hz. The parameter $r_{com}$ indicates three wireless communication ranges in the V2V environment, where the scenario $0$\,m radius indicates the improvement of local perception by receiving the estimated noise value of on-board sensors from edge. We define the improvement rate $\delta\,=\,{(MSE-MSE^{*})}/{MSE}$ as the performance metric of accuracy augmentation.

\noindent  \textbf{Distributed Estimation Analysis.} The upper part of the Tab.~\ref{table:dist_est} shows the one second average improvement rate (AvgImpRate) with respect to the ground truth $\delta_{gt,1s}$. 
As expected, the higher $f_{sub}$ and $r_{com}$ are, the faster the distributed estimation converges.
Note that the difference between $10$\,Hz and $1$\,Hz is rather small, i.e., $4-8$\,\%. Nevertheless, the results show a remarkable improvement by subscribing the noise estimation service comparing with no subscription, around $50$\,\% after $1$\,s and above $80$\,\% after $5$\,s  with a oscillation within $3$\,\%. The lower part of Tab.~\ref{table:dist_est} shows the one second average improvement rate with respect to the limit of the distributed estimation $\delta_{lim,1s}$ in each CAV, i.e., the distributed estimation with noise ground truth. After $1$\,s the distributed estimation accuracy has been improved over $90$\,\% comparing to no subscription and over $99.8$\,\% after $5$\,s. In other words, the distributed estimation by subscribing the BiFNoE based 
service is nearly the same as the estimation with true noise values.

\setlength{\tabcolsep}{3pt}
\begin{table}[t]
\begin{threeparttable}
\caption{\centering 1\,s average improvement rate (AvgImpRate) of the distributed estimation enhanced by the BiFNoE algorithm}
\label{table:dist_est}

\begin{tabular}{C{0.5cm}C{0.5cm}C{1.6cm}C{1.6cm}C{1.6cm}C{1.6cm}}
\toprule 
    
\multicolumn{2}{c}{ \textbf{Parameters}}  & \multicolumn{4}{c}{ \textbf{1\,s AvgImpRate 
to ground truth $\delta_{gt,1s}$ [\%] from $t_0$ [s]}}\\
 
{$r_{com}$} & {$f_{sub}$}  &  {$1$}  &  {$2$}  &  {$5$} &  {$10$} \\

\midrule 

 {$0$} &  {$1$}  &  {$45.73\pm12.60$} &  {$64.23\pm9.06$}  &  {$80.86\pm2.78$} &  {$85.18\pm1.35$}\\

 {$0$} &  {$10$}  &  {$53.36\pm11.31$} &  {$69.05\pm7.93$}  &  {$81.25\pm2.67$} &  {$85.25\pm1.34$}\\

 {$80$} &  {$1$}  &  {$55.23\pm11.88$} &  {$72.16\pm7.60$}  &  {$84.18\pm2.37$} &  {$85.85\pm1.83$}\\

 {$80$} &  {$10$}  &  {$59.75\pm10.08$} &  {$73.58\pm6.42$}  &  {$84.23\pm2.29$} &  {$85.87\pm1.83$}\\

 {$150$} &  {$1$}  &  {$60.60\pm11.05$}  &  {$76.57\pm4.93$}  &  {$84.22\pm2.57$} &  {$86.88\pm2.17$}\\

 {$150$} &  {$10$}  &  {$64.16\pm9.40$}  &  {$77.47\pm4.04$}  &  {$84.23\pm2.55$} &  {$86.87\pm2.18$}\\
\toprule 
    
\multicolumn{2}{c}{\textbf{Parameters}}  & \multicolumn{4}{c}{ \textbf{1\,s AvgImpRate to limit $\delta_{lim,1s}$ [\%] from $t_0$ [s]}}\\

 {$r_{com}$} &  {$f_{sub}$}  &  {$1$}  &  {$2$}  &  {$5$} &  {$10$}\\

\midrule 
 {$0$} &  {$1$}  &  {$98.28\pm2.90$}  &  {$99.28\pm0.32$}  &  {$99.87\pm0.06$} &  {$99.99\pm0.01$}\\

 {$0$} &  {$10$}  &  {$98.28\pm2.90$}  &  {$99.28\pm0.32$}  &  {$99.87\pm0.06$} &  {$99.99\pm0.00$}\\

 {$80$} &  {$1$}  &  {$92.39\pm4.40$}  &  {$98.09\pm1.70$}  &  {$99.89\pm0.03$} &  {$99.98\pm0.00$}\\

 {$80$} &  {$10$}  &  {$98.95\pm0.66$}  &  {$99.60\pm0.20$}  &  {$99.94\pm0.01$} &  {$99.99\pm0.00$}\\

 {$150$} &  {$1$}  &  {$94.12\pm4.01$}  &  {$98.35\pm1.12$}  &  {$99.89\pm0.04$} &  {$99.98\pm0.01$}\\

 {$150$} &  {$10$}  &  {$99.16\pm0.54$}  &  {$99.63\pm0.11$}  &  {$99.94\pm0.17$} &  {$99.98\pm0.00$}\\

\bottomrule 
\end{tabular}
\end{threeparttable}
\vspace{-0.2cm}
\end{table}


\subsection{CARLA-based Evaluation}

\noindent  \textbf{Experimental Design.} In order to evaluate our method in consideration of road topology, enriched traffic environment, vehicle dynamics and AI-based sensor perception models from raw data to object tracking, we demonstrate the BiFNoE and the noise estimation service in the open-source simulator CARLA~\cite{dosovitskiy2017carla} with 50\,CAVs and 50\,non-CAVs, which is similar to the scenario studied in~\cite{Shorinwa2020}. All CAVs locally
process the video in $10$\,fps and share the sensor data as object lists with the edge server and the neighbor CAVs simultaneously. We applied YOLOv3~\cite{redmon2018yolov3} to detect the objects from front and rear RGB cameras in each CAV. Additional noise is modeled in Gaussian distribution with a covariance matrix $R_i$, in which the diagonal elements are randomly set within $[0.01^2:5^2]~m^2$. With a pair of depth cameras, the position states of the detected targets are measured and fed into DKF.

\begin{figure*}[ht]
\includegraphics[trim=0 0 0 0,clip,width=0.97\linewidth]{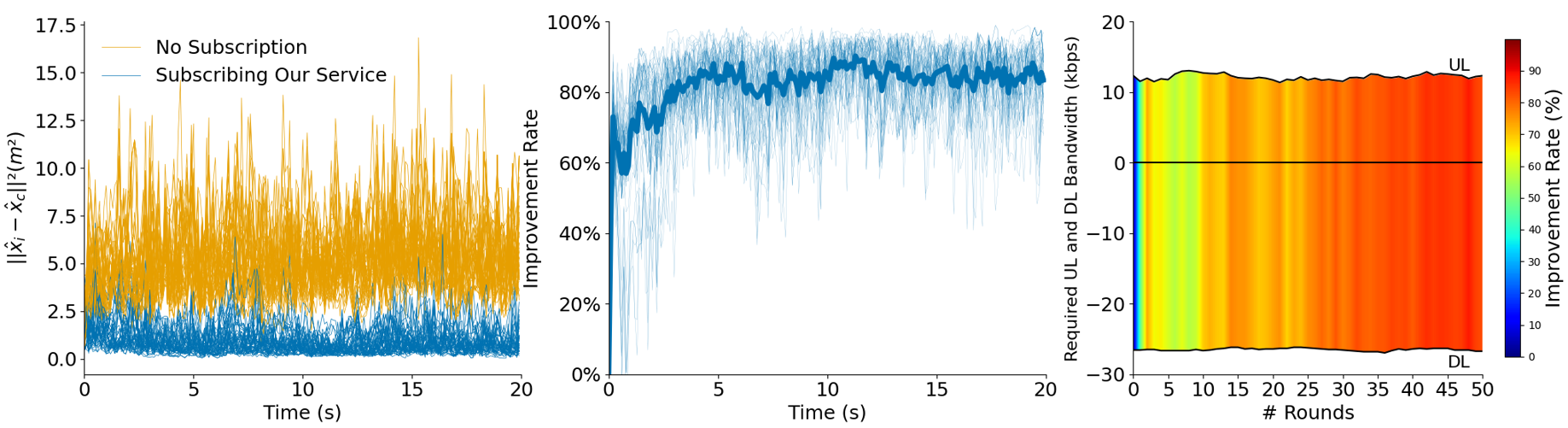}
\caption{Evaluation results from CARLA. We implemented the BiFNoE in an edge server. The CAVs in an urban scenario upload the lists of YOLOv3-detected objects through \emph{SENSORIS}~\cite{FISITA2021} to the edge server with a frequency of $10$\,Hz. Meanwhile the CAVs subscribe the noise estimation service with the same frequency. The left figure provides the MSE to the centralized estimation of each CAV with and without subscribing the proposed service. The improvement rate by subscribing the service is shown in the middle figure, where the thick line represents the average improvement rate. The right figure provides the required average uplink (UL) and downlink (DL) bandwidth per CAV in kbps (DL bandwidth is plotted at the negative area in y-axis), where the color indicates the improvement rate by subscribing the service with growing data exchange rounds.}
\label{fig:carla_sim}
\vspace{-0.3cm}
\end{figure*}

\noindent\textbf{Distributed Estimation Analysis.} Fig.~\ref{fig:carla_sim} provides the evaluation results of the CARLA simulation. The left figure shows the MSE to centralized estimation of each CAV with and without our proposed service. The errors of distributed estimations in all CAVs are clearly reduced by subscribing the noise estimation service. The middle figure further indicates the improvement rate in each CAV as well as the on average values plotted with the thick line for all CAVs. After about $3$\,s, the distributed estimations with subscription are on average improved by around $80\%$, which is consistent with the simulation result in Sec.~\ref{sec:system}.  The right figure depicts the required communication cost for our service in each CAV. Through transmitting the \emph{SENSORIS} messages~\cite{FISITA2021} for noise estimation service, only around $12$\,kbps uplink and $28$\,kbps downlink bandwidth for each CAV are required for uploading the sensor data to edge server and subscribing the noise values. The improvement rate converges to $80$\,\% after only $30$ rounds of data exchange between CAVs and the edge server.

\section{Conclusion}
\label{sec:conclusion}


This paper addresses the problem of unknown and time-varying sensor noise values in vehicular communication networks, which can cause failures in the distributed estimation as part of the sensor data sharing and fusion for sensor data sharing in support of connected automated vehicles. The proposed algorithm BiFNoE accomplishes the noise estimation in an edge server by receiving the shared sensor data in a dedicated vehicle fleet. It can provide reliable estimated measurement noise of each CAV, and thereby notably increase the accuracy of centralized 
target estimation. The evaluation results in the simulation environment using the autonomous driving simulator CARLA indicate that the distributed estimations at the associated CAVs are improved by around $80$\,\% on average with small communication UL and DL bandwidth for subscribing the proposed noise estimation service. Future work will aim at investigating the performance of the algorithms in more complex traffic scenarios and an unstable communication environment in the context of realistic vehicular data traffic.


\bibliography{ref}
\bibliographystyle{IEEEtran}


\end{document}